\documentclass[twoside,prb,letterpaper,preprintnumbers,superscriptaddress,notitlepage,onecolumn]{revtex4}

\usepackage{mathptmx}
\usepackage{helvet}
\usepackage{graphicx,fancyhdr}% Include figure files
\usepackage{dashbox,color}%
\usepackage{bm}% bold math
\usepackage{sidecap}%
\definecolor{lightblue}{rgb}{0.6,0.9,1}
\definecolor{myrefblue}{rgb}{0.1,0.6,1}
\definecolor{myblue}{rgb}{0,0,0}
\definecolor{nmat}{rgb}{0.7,0.04,0.26}

\begin{document}
\pagestyle{fancy}

\renewcommand{\headrule}{\vskip-3pt\hrule width\headwidth height\headrulewidth \vskip-\headrulewidth}

\fancypagestyle{plainfancy}{%
\lhead{}
\rhead{}
\chead{}
\lfoot{}
\cfoot{}
\rfoot{\bf\scriptsize\textsf{\thepage}}
\renewcommand{\headrulewidth}{0pt}
\renewcommand{\footrulewidth}{0pt}
}

\fancyhead[LE,RO]{}
\fancyhead[LO,RE]{}
\fancyhead[C]{}
\fancyfoot[LO,RE]{}
\fancyfoot[C]{}
\fancyfoot[LE,RO]{\bf\scriptsize\textsf{\thepage}}

\renewcommand\bibsection{\section*{\sffamily\bfseries\normalsize {References}\vspace{-10pt}\hfill~}}
\newcommand{\mysection}[1]{\section*{\sffamily\bfseries\normalsize {#1}\vspace{-10pt}\hfill~}}
\renewcommand{\subsection}[1]{\noindent{\bfseries\normalsize #1}}
\renewcommand{\bibfont}{\fontfamily{ptm}\footnotesize\selectfont}
\renewcommand{\figurename}{Figure}
\renewcommand{\refname}{References}
\renewcommand{\bibnumfmt}[1]{#1.}

\newcommand{\be}{\begin{equation}}
\newcommand{\ee}{\end{equation}}
\newcommand{\bn}{\begin{eqnarray}}
\newcommand{\en}{\end{eqnarray}}
\newcommand{\ii}{\'{\i}}
\newcommand{\ca}{\c c\~a}

\makeatletter
\long\def\@makecaption#1#2{%
  \par
  \vskip\abovecaptionskip
  \begingroup
   \small\rmfamily
   \sbox\@tempboxa{%
    \let\\\heading@cr
    \textbf{#1\hskip1pt$|$\hskip1pt} #2%
   }%
   \@ifdim{\wd\@tempboxa >\hsize}{%
    \begingroup
     \samepage
     \flushing
     \let\footnote\@footnotemark@gobble
     \textbf{#1\hskip1pt$|$\hskip1pt} #2\par
    \endgroup
   }{%
     \global \@minipagefalse
     \hb@xt@\hsize{\hfil\unhbox\@tempboxa\hfil}%
   }%
  \endgroup
  \vskip\belowcaptionskip
}%
\makeatother

\thispagestyle{plainfancy}

\fontfamily{helvet}\fontseries{bf}\selectfont
\mathversion{bold}
\begin{widetext}
\begin{figure}
\vskip0pt\noindent\hskip-0pt
\hrule width\headwidth height\headrulewidth \vskip-\headrulewidth
\hbox{}\vspace{4pt}
\hbox{}\noindent\vskip10pt\hbox{\noindent\huge\sffamily\textbf{Microscopic description of insulator-metal transition}}\vskip0.05in\hbox{\noindent\huge\sffamily\textbf{in high-pressure oxygen}}
\vskip10pt
\hbox{}\noindent\begin{minipage}{\textwidth}\flushleft
\renewcommand{\baselinestretch}{1.2}
\noindent\hskip-10pt\large\sffamily Luis Craco$^{1}$, Mukul S. Laad$^2$ \& 
Stefano Leoni$^{3}$
\end{minipage}
\end{figure}
\end{widetext}

\begin{figure}[!h]
\begin{flushleft}
{\footnotesize\sffamily
$^1$Instituto de F\ii sica, Universidade Federal de Mato Grosso, 78060-900, 
Cuiab\'a, MT, Brazil. $^2$The Institute of Mathematical Sciences, C.I.T. 
Campus, Chennai 600 113, India. $^3$School of Chemistry, Cardiff University, 
Cardiff, CF10 3AT, UK.}
\end{flushleft}
\end{figure}

%\small
\fontsize{9pt}{8pt}\selectfont
\renewcommand{\baselinestretch}{0.9}
%abstract
\noindent\sffamily\bfseries{Unusual metallic states involving breakdown of 
the standard  Fermi-liquid picture of long-lived quasiparticles in 
well-defined band states emerge at low temperatures near correlation-driven 
Mott transitions. Prominent examples are ill-understood metallic states in 
$d$- and $f$-band compounds near Mott-like transitions. Finding of 
superconductivity in solid O$_{2}$ on the border of an insulator-metal 
transition at high pressures close to 96~GPa is thus truly remarkable. 
Neither the insulator-metal transition nor superconductivity are understood 
satisfactorily. Here, we undertake a first step in this direction by focussing 
on the pressure-driven insulator-metal transition using a combination of 
first-principles density-functional and many-body calculations. We report a 
striking result: the finding of an orbital-selective Mott transition in a pure 
$p$-band elemental system. We apply our theory to understand extant structural 
and transport data across the transition, and make a specific two-fluid 
prediction that is open to future test. Based thereupon, we propose a novel 
scenario where soft multiband modes built from microscopically coexisting 
itinerant and localized electronic states are natural candidates for the 
pairing glue in pressurized O$_{2}$.}

\mathversion{normal}
\normalfont\normalsize

\section{Introduction}

The unique properties of high-pressure induced solid phases of molecular 
gases continue to evince keen and enduring interest in condensed matter 
physics. Beginning with early ideas of Mott~\cite{[mott]} and extending 
up to modern times~\cite{[ashcroft]}, ideas of pressure-induced 
electronic, magnetic and structural transitions and possible 
superconductivity in such systems even provided early ground for strongly 
correlated systems, are currently a frontline research topic in condensed 
matter. Particularly interesting examples of intriguing physics in solidized 
molecular phases of gases are dense hydrogen~\cite{hydro} and solid 
oxygen~\cite{solox,sol}, as well as the most recent report of very high-$T_{c}$ 
superconductivity in solid H$_{2}$S under very high pressure~\cite{eremets}. 
H$_{2}$ is predicted to metallize under high pressure, while solid O$_{2}$ 
even shows a superconducting phase ($T_{c}=0.6$~K) at the border of a 
pressure-driven transition from a non-magnetic insulator to paramagnetic 
metal, joining the long list of materials exhibiting superconductivity 
proximate to metal-insulator transitions.

Pressurized molecular oxygen forms various low-temperature solid phases
under pressure, labelled $\alpha$, $\delta$, $\epsilon$ and $\zeta$ 
phases~\cite{kat}. At lower pressure, the antiferromagnetically ordered 
$\alpha$ phase transforms into another antiferromagnetically ordered 
$\delta$ phase at $5.4$~GPa, followed by a non-magnetic $\epsilon$ phase 
at $8$~GPa. Higher pressure, $P\simeq 96$~GPa, metallizes the 
system~\cite{metal}, followed by emergence of superconductivity below 
$T_{c}\simeq 0.6$~K~\cite{super}. This astounding behavior in a molecular 
system, reminiscent of strongly correlated, doped Mott insulators in 
$d$-band oxides like cuprates, presents a significant challenge for 
theory. The high-$P$ $\epsilon-\zeta$ phase transition is also accompanied 
by significant volume reduction~\cite{weck}, with a contraction of about 
$10\%$ of the lattice parameter along the $b$ direction. The $\epsilon$ phase
retains the layered nature of the lower pressure phases~\cite{solox},
and the monoclinic ($C2/m)$ structure~\cite{weck,fuji} as shown in 
Fig.~\ref{fig0}.

\begin{figure}[!b]
\vskip50pt
\end{figure}
\begin{figure*}[!t]
\includegraphics[width=5.1in]{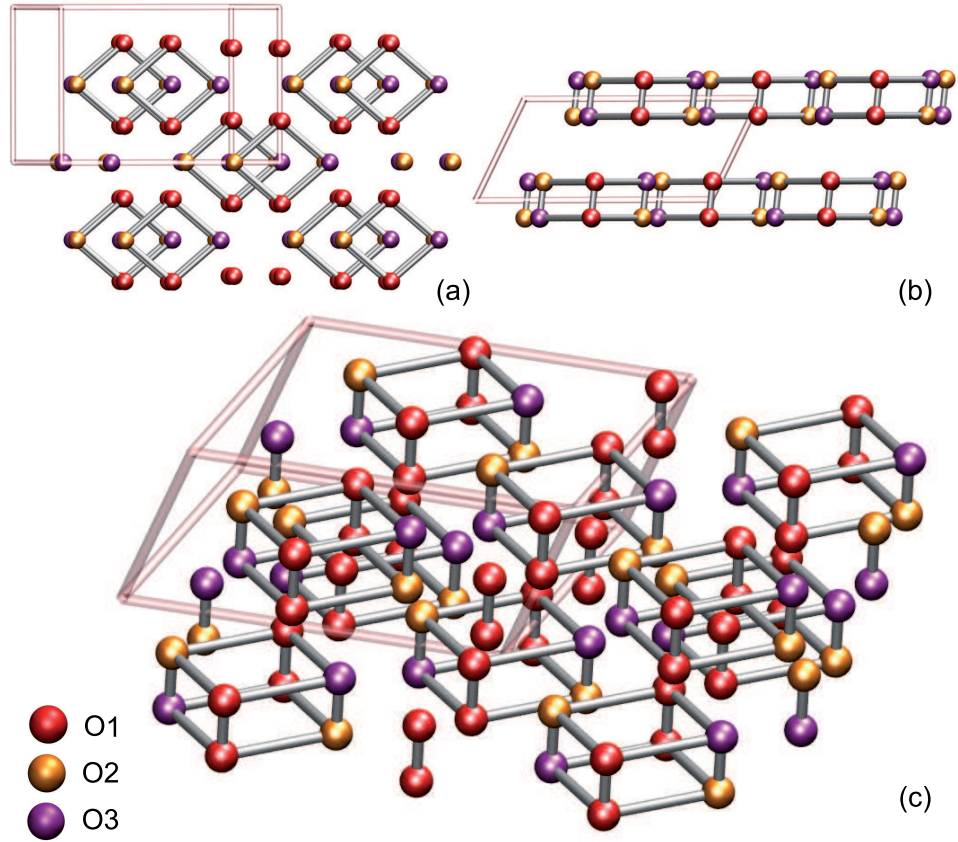}
\caption{{\bf Crystal structure of the $\epsilon$-phase of solid oxygen.}
The structure as viewed perpendicular to the ${\bf a}-{\bf b}$ (a) and 
${\bf a}-{\bf c}$ (b) planes. The O$_8$ clusters in the monoclinic 
unit cell (rose lines) are shown in (c). O$x (x=1,2,3)$ label the 
three inequivalent oxygen atoms.}
\label{fig0}
\end{figure*}

That the driving force for the $\alpha-\beta$ transition at moderate $T$
is dominantly magnetic has been established in a series of careful
studies~\cite{ab,ab1,ab2}. Indeed, early work of da Silva and 
Falicov~\cite{[leo]} already explained the measured heat of formation at 
the $\alpha-\beta$ transition in terms of the entropy difference computed 
from cluster analysis of a multi-orbital Hubbard model (or an equivalent 
$S=1$ Heisenberg-like model in $d=2$ dimensions). Observation of very 
different magnetic orders in the $\alpha,\beta$ phases, correlation
between magnetic and structural changes along with ferromagnetic coupling 
between the off-plane near neighbors in the $\delta$ phase are reminiscent 
of those found in classic multi-band systems like V$_{2}$O$_{3}$~\cite{[bao]}, 
taken together with the above, favor a multi-orbital description. Additional 
evidence for multi-orbital effects is provided by the anisotropic and 
partially discontinuous pressure-induced changes in the lattice parameters 
in the different phases~\cite{fuji,ma}. In such a scenario, increasing 
pressure is expected, in the simplest approximation, to decrease lattice 
spacings and increase the carrier itinerance. The result would then be to 
suppress antiferromagnetic order along with insulating behavior, and to 
induce metalization. In solid O$_{2}$, antiferromagnetic order is destroyed 
well before metalization occurs~\cite{goncha}, and so, within the $p^{4}$ 
configuration of oxygen, the insulator-metal transition across the 
$\epsilon-\zeta$ transition must be regarded as a Mott metal-insulator 
transition. This suggests that on the one extreme, a Heisenberg 
model description is only valid in the insulating $\alpha,\beta,\delta$ 
phases, and that a more general multi-orbital Hubbard model must be used, 
at least for the $\epsilon$ phase. At the other extreme, one-electron 
band structure calculations for the  antiferromagnetically ordered phases 
do provide qualitatively correct ground states~\cite{serra}. In addition, 
electronic structure calculation based on generalized gradient 
approximation (GGA) shows that the nonmagnetic insulating state is 
energetically favored at pressures corresponding to the 
$\epsilon$-phase~\cite{ashcroft,calle}. However, by construction, 
{\it ab initio} density-funtional calculations have intrinsic 
difficulties in describing non-magnetic insulating phases, and 
in particular the $\epsilon$ phase~\cite{tse,barto}, for reasons described 
in detail in ref.~\onlinecite{cohen}. The observation of superconductivity  
at the border of this (Mott) insulator-to-metal transition thus suggests that 
dualistic behavior of correlated carriers (Mottness) near the insulator-metal 
transition is very likely implicated in the pairing glue. Thus, a search for 
the microscopic origin of the pair glue must involve understanding of the 
insulator-metal transition around $96$~GPa.

\begin{figure}[!b]
\vskip50pt
\end{figure}
\begin{figure*}[!t]
\includegraphics[width=6.1in]{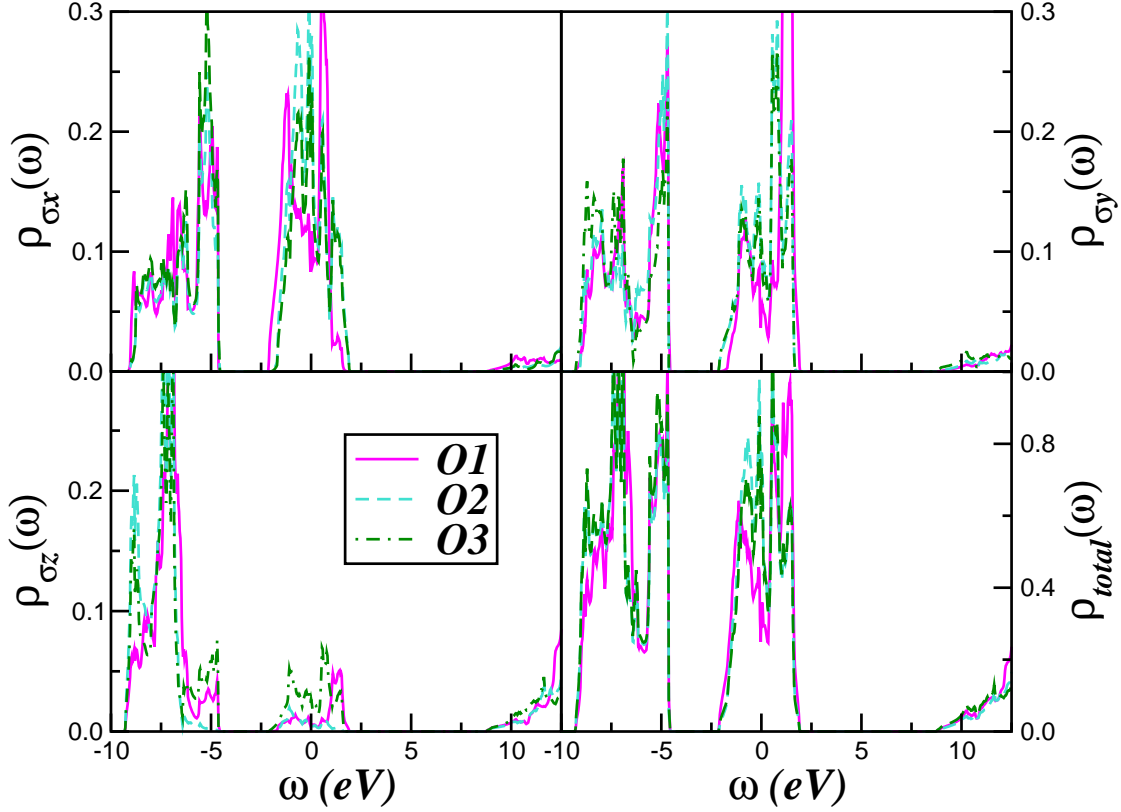}
\caption{{\bf Orbital resolved and total LDA density-of-states (DOS) 
for the three inequivalent oxygen atoms in the $\epsilon$-phase.}
Notice that all bands as well span over the Fermi level. Also 
relevante is the evolution of the electronic DOS at different 
polarizations.}
\label{fig1}
\end{figure*}

Before presenting our local-density-approximation plus dynamical-mean-field 
(LDA+DMFT) results, we point out essential differences between band and Mott 
insulators. In conventional semiconductors (or band insulators) all bands 
below the Fermi energy are filled and, therefore, inert. Removing an electron 
leads to an empty state which can be thought of as a hole moving freely 
through the solid. The same is true for an added electron, which occupies 
the first empty band. In a multi-orbital Mott-Hubbard insulator, the 
insulating state arises because electron hopping from one site to another 
is inhibited by intra- and inter-orbital Coulomb repulsions. In these systems, 
when the band filling is slightly reduced from its commensurate value, a small 
number of unoccupied states are created; similarly adding electrons creates 
locally doubly occupied electronic states. The crucial difference in this case
is that since the doped carriers can have either spin ($\uparrow,\downarrow$) 
with equal probability, doping a Mott insulator, $e.g$, by holes, creates 
two available states at the Fermi energy.  This is at the heart of spectral 
weight transfer, a phenomenon ubiquitous to Mott, as opposed to band, 
insulators. In both cases, electron hopping might still be prevented by 
inter-orbital Coulomb interactions in a multiband system. The resulting 
metallic state upon doping can vary from a Fermi liquid at weak coupling 
to an exotic orbital-selective, non-Fermi liquid metal for stronger 
electron-electron interactions, as doping and temperature~\cite{nFL} are 
varied. This fundamental difference between band and multi-orbital 
Mott-Hubbard insulators is of basic and practical interest. Below we show 
that sizable multiband electronic interactions are the clue to the insulating 
state of the $\epsilon$-phase of solid oxygen and its evolution to a 
non-Fermi liquid metallic state at high pressures.

Possibility of Mott-Hubbard physics in purely 
$p$~\cite{zunger,zun1,zun2,strained} or $s$~\cite{chiape} band systems is 
very intriguing, since the naive expectation dictates that the itinerance 
(kinetic energy of $p,s$-carriers) is appreciable compared to the 
electron-electron interactions, as distinct from $d$-band systems, where 
the $d$ electrons reside in much narrower bands (hence the effective $U/W$ 
is sizable; $U$ and $W$ are, respectively, the on-site Coulomb repulsion 
and the bare one-particle band width)~\cite{kotrev}. Thus, understanding 
Mottness in solidified gases with active $p$ or $s$ bands is undoubtedly 
an issue of great contemporary interest. In light of the discussion above, 
we study how an orbital-selective interplay between appreciable $p$-band 
itinerance and sizable, on-site Coulomb repulsion, $U$, plays a central 
role in this unique Mott transition in solid O$_2$.

\section{Results}

\subsection{Electronic Structure}

To quantify the correlated electronic structure of solid O$_2$, we start 
with the $C2/m$ structure (Fig.~\ref{fig0}) with lattice parameters 
derived in ref.~\onlinecite{fuji}. Here, local-density approximation 
(LDA) calculations for the real crystal structure of the $\epsilon$-phase 
were performed using the linear muffin-tin orbitals (LMTO)~\cite{ok,claudia} 
scheme in the atomic sphere approximation~\cite{sphere}. The corresponding 
LDA density-of-states of the three (symmetry) inequivalent 
atoms~\cite{solox,fuji} is shown in Fig.~\ref{fig1}. Strong intramolecular 
overlap leads to propensity to localization of the $p_z$, i.e, the 
$\sigma$-orbital~\cite{serra,meng} in the energy level diagram of O$_2$. 
However, due to inter-molecular orbital overlap in the monoclinic structure, 
the $p_{z}$ states acquire some itinerance, explaining the small amount of 
$p_{z}$ states found at the Fermi energy. As seen in Fig.~\ref{fig1}, all 
$\pi$-bands cross the Fermi energy, providing a metallic state within LDA. 

Within LDA, the one-electron part of the many-body Hamiltonian for solid 
oxygen is now $H_{0} =\sum_{{\bf k},a,\sigma}
\epsilon_{a}({\bf k})c_{{\bf k},a,\sigma}^{\dag}c_{{\bf k},a,\sigma}  
+ \sum_{i,a,\sigma}\Delta_{a}n_{i\sigma}^{a}$,
where $a=x,y,z$ label the three diagonalized $p$ orbitals and the $\Delta_{a}$ 
are on-site orbital energies in the real structure of solid O$_{2}$. In light 
of antiferromagnetic insulator~\cite{[leo]} phases and the non-magnetic Mott 
transition, local multi-orbtial interactions are mandatory to understand 
O$_{2}$. These constitute the interaction terms 
$H_{int}=U\sum_{i,a} n_{i\uparrow}^{a} n_{i\downarrow}^{a}
+ U'\sum_{i,a \ne b}n_{i}^{a}n_{i}^{b}
-J_{H}\sum_{i,a \ne b}{\bf S}_{ia} \cdot {\bf S}_{ib}$.
Here, $U~(U'\equiv U-2J_H)$ is the intra- (inter-) orbital Coulomb repulsion 
and $J_H$ is the Hund's rule term. Following da Silva and Falicov~\cite{[leo]}, 
we use $U=11.6$~eV and $J_H=0.45$~eV, along with the LDA bands of the three 
inequivalent oxygen atoms described above. In this work, the correlated 
multi-orbital problem of solid O$_2$ encoded in $H=H_{0}+H_{int}$ is treated 
within the state-of-the-art local-density-approximation plus 
dynamical-mean-field-theory (LDA+DMFT) scheme~\cite{kot-rev}. The DMFT 
self-energy, $\Sigma_{a}(\omega)$, requires a solution of the multi-orbital 
quantum impurity problem self-consistently embedded in an effective 
medium~\cite{kot-rev}. We use the multi-orbital iterated-perturbation-theory 
(MO-IPT) as an impurity solver for DMFT~\cite{ePAM}: This analytic solver has 
a proven record of successes in describing finite temperature Mott 
transitions~\cite{v2o3} as well as unconventional behavior in correlated 
$p$-band systems~\cite{ourbi,bite,bism}.

\begin{figure}[!b]
\vskip50pt
\end{figure}
\begin{figure*}[!t]
\includegraphics[width=6.1in]{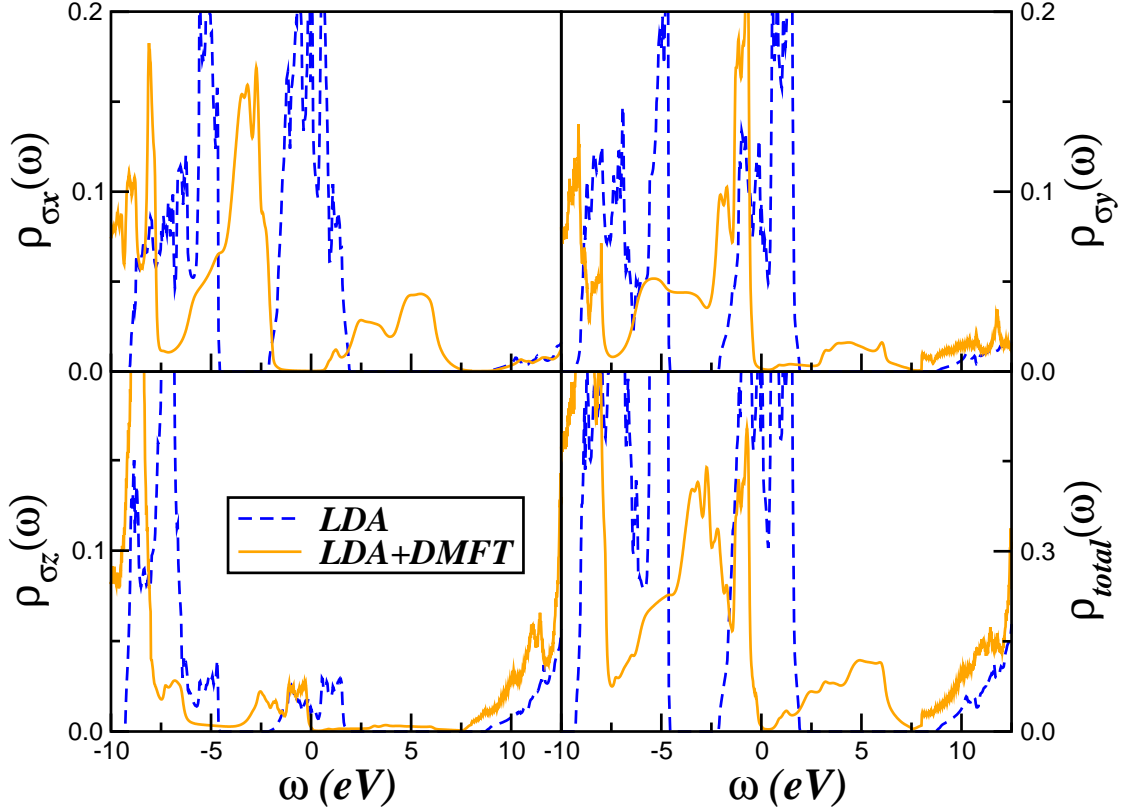}
\caption{{\bf Comparison between the LDA and LDA+DMFT orbital-resolved 
and total density-of-states in the $\epsilon$-phase of solid O$_2$.} 
LDA+DMFT results for the Mott insulating $\epsilon$ phase of oxygen were 
obtained using the intra- (inter-) orbital Coulomb repulsion, $U=11.6$~eV 
($U'=10.7$~eV) and the Hund's rule interaction  $J_H=0.45$~eV for the total 
band filling, $n$=4. Large-scale transfer of spectral weight from low energy 
to high energies is visible in the correlated spectral functions of the $p_x$ 
and $p_y$ bands. Also clear is the destruction of the low-energy peak of the 
$p_z$ in LDA.}
\label{fig2}
\end{figure*}

With orbital orientation-induced anisotropic LDA one-particle energies and 
hoppings, multi-orbital correlations renormalize various $p$-bands in 
different ways. Generically, one expects partial (Mott) localization of a 
subset of bands, leading to orbitally selective Mott transitions, and bad 
metallic states~\cite{kotrev,v2o3,ytio}. Within LDA+DMFT, this 
orbital-selective mechanism involves two renormalizations: static 
(multi-orbital Hartree) renormalization shifts the $p$-bands relative to 
each other by amounts depending upon their {\it bare} on-site orbital 
energies ($\Delta_a)$ and occupations ($n^a$). In addition, dynamical 
effects of $U,U'\equiv U-2J_H$ drive large spectral weight transfer over 
wide energy scales~\cite{v2o3,ytio}. The large, anisotropic changes in 
dynamical spectral weight transfer in response to small changes in bare 
one-particle (LDA) parameters (for example, crystal-field splittings under 
pressure)~\cite{v2o3} are known to drive the orbital-selective Mott 
transition in real multi-orbital systems. As we show below, precisely 
such an orbital-selective Mott transition, accompanied by an incoherent 
metallic phase in solid O$_2$, occurs at very high pressures.

Using $U,U',J_{H}$ as obtained in ref.~\onlinecite{[leo]}, we find that 
the $\epsilon$ phase is a Mott insulator, as shown in Fig.~\ref{fig2}. 
The size of the charge gap is orbital dependent, and is larger for 
the $p_{x}$ band compared to the other two. Large spectral weight 
transfer, characteristic of dynamical local correlations, is explicitly 
manifested in the qualitative difference between LDA and LDA+DMFT spectra.
 At first sight, derivation of a Mott insulator with $U<W$ in 
solid O$_{2}$ seems a bit puzzling. The reason, however, is that, 
in this multi-orbital system, both $U,U'$ are appreciable, 
and the combined effect of both acting in tandem is to ($i$) reduce 
the band-width of each band (this can arise solely from $U$, even 
for the artificial case of $U'=0$), and ($ii$) the dominant effect 
of $U'$ on a reduced bandwidth is to split the bands via the Mott
mechanism. In the actual multi-orbital problem, both effects are 
simultaneously operative, and reinforce each other. 

From our results in Fig.~\ref{fig2}, we compute the renormalized orbital 
splittings ($\delta_a$) and occupations ($n^a$) within LDA and LDA+DMFT. 
Within LDA, we find $(\delta_x,\delta_y,\delta_z)=(-1.12,-0.95,-5.22)$~eV 
and $(n^{x}_{\sigma},n^{y}_{\sigma},n^{z}_{\sigma}=0.68,0.59,0.53)$. LDA+DMFT 
severely renormalizes the center of gravity of each band to 
$(\delta_x,\delta_y,\delta_z)=(-3.76,-2.93,-6.65)$~eV, as well as the 
orbital occupancies to 
$(n^{x}_{\sigma},n^{y}_{\sigma},n^{z}_{\sigma}=0.69,0.78,0.56)$,
promoting enhanced orbital polarization. This fact, generic to 
multi-orbital systems (though we do not find total orbital
polarization)~\cite{liebsch}, is an interesting manifestation of
correlation-induced orbital rearrangement, and controls structural
changes across the Mott transition (see below).

\begin{figure}[!b]
\vskip50pt
\end{figure}
\begin{figure*}[!t]
\includegraphics[width=6.1in]{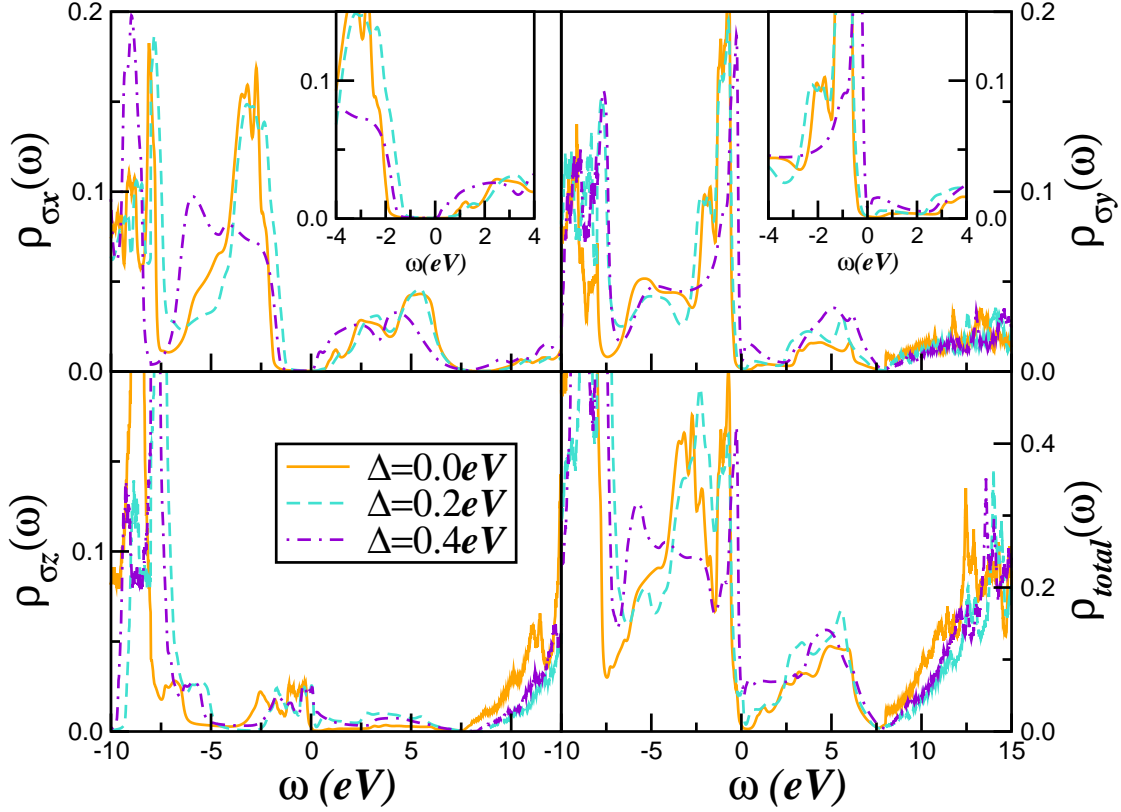}
\caption{{\bf Orbital-selective insulator-to-metal transition in 
pressurized O$_2$.} In our theory, we vary the trial orbital-splitting 
$\Delta$ within LDA+DMFT to simulate structural changes upon pressure. 
In our results for the orbital-selective metallic phase the $p_y$ 
(and hence, total) DOS shows a clear pseudogap around $E_F$, 
corresponding to an orbital-selective, non-Fermi liquid metallic phase. 
Inset at the top panels show the evolution of the electronic states 
close to the Fermi energy.}
\label{fig3}
\end{figure*}

We now turn to the insulator-metal transition in solid O$_2$ at high $P$, 
and adopt the following strategy to derive this transition. Instead of 
reverting back to the LDA to use a different LDA density-of-states 
corresponding to the metallic $\zeta$-phase, we search for an instability 
of the insulating, $\epsilon$ phase to the paramagnetic-metal by varying 
$\delta_a$, found for the Mott insulator above. To proceed, consider the 
orbital-dependent on-site energy term, 
$H_{\Delta}=\sum_{i,a,\sigma}\Delta_{a} n_{i\sigma}^{a}$ in our Hamiltonian. 
We now let the trial $\Delta$ vary in small steps, keeping 
$\Delta_{x}=-\Delta$, $\Delta_{y}=\Delta$, to simulate the structural
(and hence, electronic) changes upon pressure. As for V$_{2}$O$_{3}$~\cite{v2o3} 
and YTiO$_3$~\cite{ytio}, we search for the second self-consistent LDA+DMFT 
solution by solving the multi-orbital DMFT equations for each trial value of 
$\Delta$ keeping $U,U'$ fixed. As seen in Fig.~\ref{fig3}, small variations of 
$\Delta$ drive appreciable spectral weight transfer, producing drastic 
orbital-selective renormalizations of the one-particle spectral functions: the 
$p_x$ and $p_y$ are most severely affected. At a critical $\Delta_{c}=0.3$~eV, 
the $p_x$ density-of-states remains Mott insulating, while the $p_y$ band 
undergoes an insulator to bad-metal (weakly first-order, with no coherent 
Kondo peak at $E_F$) transition. Thus, our results imply that the paramagnetic, 
metallic phase of $\zeta$-oxygen is an orbital-selective incoherent metal 
without Landau quasiparticles, characterized by a pseudogap at $E_F$ in the 
$p_y$, and hence, in the total spectral function, at $E_F$. Our simulations 
(not shown) indicate that Kondo-like resonance found below $E_F$ for 
$\Delta=0.4$~eV will cross the Fermi level at extremely high pressures, 
driving solid O$_2$ into (quasi)coherent Fermi liquid-like metallic state 
at even higher (experimentally uninvestigated) pressures. The underlying 
theoretical reason for this is as follows: At $\Delta_c$, strong scattering 
between the effectively (Mott) localized and itinerant components of the 
matrix DMFT propagators produces an incoherent metal because strong interband 
scattering indeed operates in a sizably orbitally polarized metallic system.  
However, at very high pressure (large $\Delta > \Delta_c$), the $p_x$ band 
becomes almost fully polarized (Fig.~\ref{fig6}, upper panel) and the system 
evolves into a low-$T$ correlated Fermi liquid metal~\cite{ytio}, which we 
predict to be the post-$\zeta$ phase. This is consistent with the fact that 
strong crystal-field splitting supresses local orbital fluctuations and cuts 
off the strong scattering channel. In turn, this controls the orbital-selective 
phase boundary of correlated multi-orbital systems~\cite{medici}. From 
our results, the orbital-selective Mott phase is thereby suppressed at high 
pressure, leading to continuous evolution of the incoherent, bad-metal 
to a correlated Fermi liquid like metal. 

\begin{figure}[!b]
\vskip50pt
\end{figure}
\begin{figure*}[!t]
\includegraphics[width=6.1in]{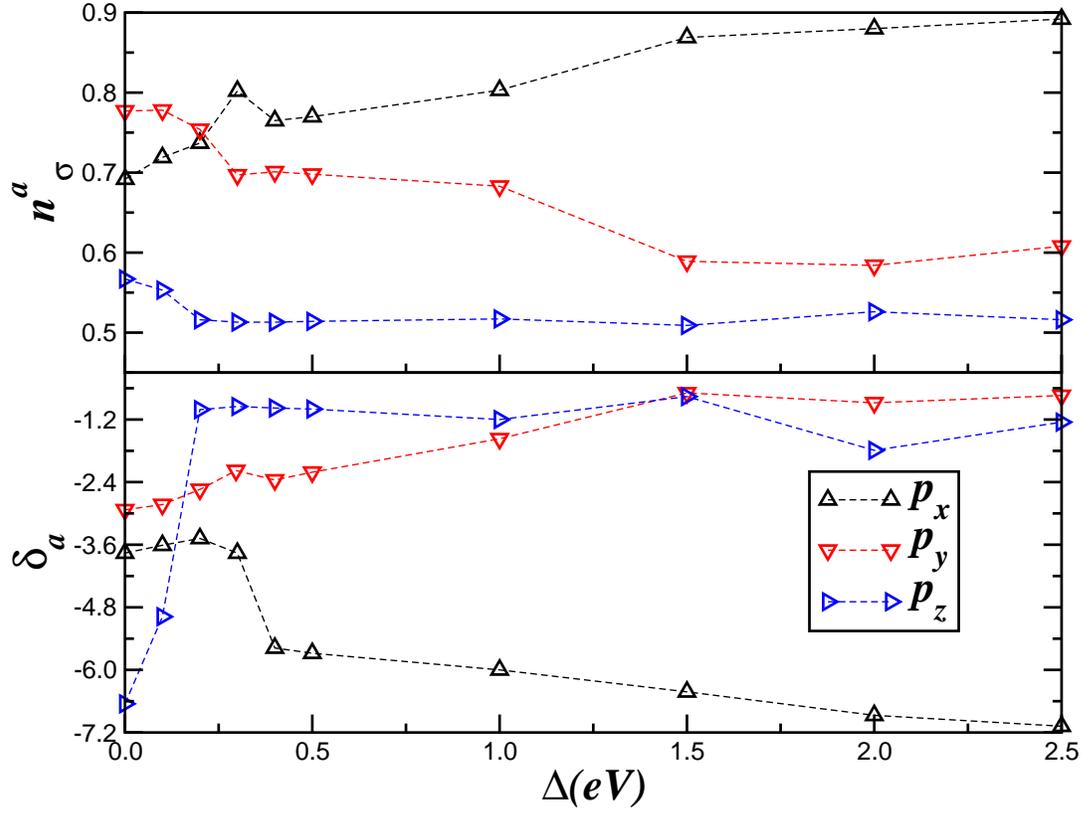}
\caption{{\bf Effect of the orbital ``Zeeman'' field upon pressure.}
(Top) LDA+DMFT results for the orbital occupations $n^{a}_{\sigma}$ and 
(bottom) the renormalized orbital splittings $\delta_{a}$. Notice that 
$n^{y}_{\sigma}$ jumps at the insulator-metal transition, a behavior 
characteristic of orbital-selective Mott transition.}
\label{fig6}
\end{figure*}

In Fig.~\ref{fig6} we show the evolution of the orbital occupations 
$n^{a}_{\sigma}$ (top) and the renormalized orbital splittings $\delta_{a}$ 
across the insulator-metal transition. The features are well understood 
as follows. In a multiband situation, $\Delta$ acts like an external 
``Zeeman'' field~\cite{v2o3,ytio} in the orbital sector. The insulator-metal 
transition is characterized by a sudden jump in the renormalized 
$\delta_{z}$, and in the $p_x$ and $p_y$ populations as a consequence, 
suggesting that anisotropic structural (and volume) changes will accompany 
the orbital-selective Mott transtion. Here, we propose that these changes 
in $n^a$ control anisotropic changes in lattice parameters ({\bf a,b,c}) 
across the insulator-metal transition: indeed, the changes in {\bf a,b,c} 
are expressible in terms of $n^{a}$ as 
$\gamma_{a}=\Delta l^{a}/l^{a}=(\frac{g}{Mv_{sa}^{2}})\Delta n^{a}$, where $g$ 
is the electron-phonon coupling constant, $M$ the ion mass, and $v_{sa}$ is 
the velocity of sound along $a~(\equiv x,y,z)$. Changes in $\gamma_{a}$ across 
the insulator-to-metal transition thus follow those in the $n^{a}$. Though 
values of $v_{sa}$ and $g$ in the $\epsilon$ phase are unknown, we deduce 
that the lattice parameter {\bf a} increases, while {\bf b,c} descrease 
across the orbital-selective Mott transition as in Fig.~\ref{fig6} upper 
panel. The correct trend vis-a-vis experiment~\cite{weck} provides further 
support for our Mottness scenario in solid O$_2$.

\vspace{0.25cm}
\subsection{Normal state resistivity}

\begin{figure}[!b]
\vskip50pt
\end{figure}
\begin{figure*}[!t]
\includegraphics[width=6.1in]{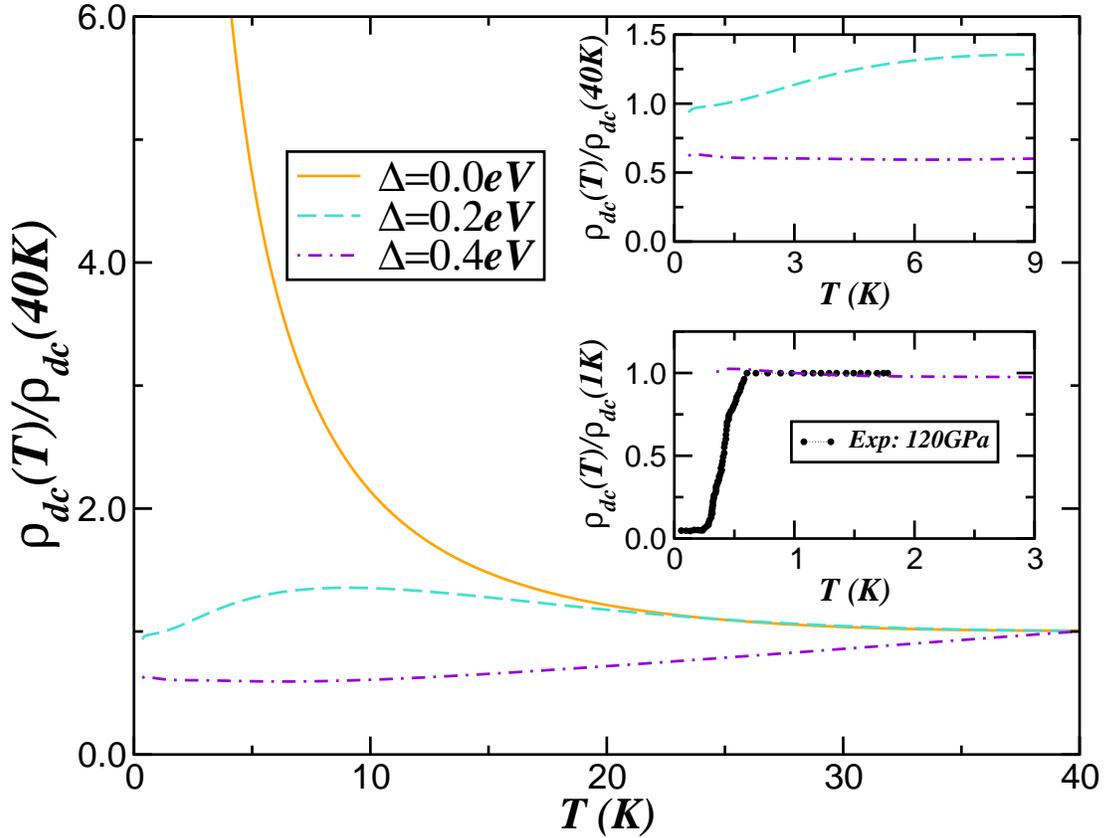}
\caption{{\bf Electrical resistivity of solid O$_2$ at high pressures.} 
Main panel: Normal state resistivity versus temperature (normalized to 
$\rho(40K)$), showing the metal-insulator transition with increasing the 
orbital ``Zeeman'' field $\Delta$. Inset shows resistivity at very low 
temperatures: Observed features at low-$T$ are well reproduced by strong 
Coulomb correlations $U,U'$ and $\Delta=0.4$~eV. Notice that results with 
small $\Delta$ value deviates from observation of constant $\rho(T)$ at 
120~GPa in experiment~\cite{super}. It is possible that a detailed experimental 
study of $\rho_{dc}(T)$ in the pressure range closer to the insulator-to-metal 
transition may reveal the trend we find, and this would constitute more 
concrete support for our modelling.}
\label{fig7}
\end{figure*}

To illustrate the importance of correlation-induced changes in the orbital 
``Zemman'' field $\Delta$ under high pressure in our theory, we now discuss 
our results for the normal state resistivity computed within the Kubo 
formalism~\cite{class}. In our theory, the observed features in $\rho_{dc}(T)$ 
originate from  changes in the correlated spectral functions with $\Delta$.  
Showing how this provides a compelling description of the admittedly limited 
available data is our focus in what follows. In Fig.~\ref{fig7}, we show the 
$\rho_{dc}(T)$ for three values of $\Delta$ in solid O$_2$, computed using the 
LDA+DMFT orbital resolved spectral functions (with $U=11.3$~eV, $U'=U-2J_H$, 
and $J_H=0.45$~eV). Various interesting features immediately stand out. First, 
$\rho_{dc}(T\rightarrow 0)$ in the $\epsilon$ phase ($\Delta=0$) shows 
semiconducting behavior, in accord with the insulating classification at lower 
pressures, when $\Delta<\Delta_{c}$.  Secondly, at all $T$, no Fermi liquid 
$T^{2}$-like contribution is detectable in the metallic phase with 
$\Delta=0.4$~eV: instead, $\rho_{dc}(T)$ is approximately constant up to 10~K.  
For intermediate pressure (but on the metallic side), $\rho_{dc}(T)$ crosses 
over from semiconductor-like (at high $T$) to bad-metallic (at low $T$) 
behavior. Remarkably, the detailed $T$-dependence closely resembles that 
seen in experiment~\cite{super}, in the normal state up to 2~K (see inset of 
Fig.~\ref{fig7}).  Since the system is proximate to a Mott transition, the 
$T$-dependence of $\rho_{dc}(T)$ for both values of $\Delta=0.2,0.4$~eV is 
characteristic for carriers scattering off dynamically fluctuating and coupled, 
short-range spin and charge correlations. On general grounds, we expect this 
effect to be relevant near a correlation-driven Mott transition.  Since an 
external magnetic field will generically quench spin fluctuations, we predict 
that destroying the $\epsilon$ Mott insulating state by a magnetic 
field~\cite{nomura} might reveal this behavior. We emphasise that resistivity 
measurements as a function of pressure over extended $T$ scales is a smoking 
gun for our proposal, as would be the study of the $T$-dependence of the $dc$ 
Hall constant. These can distinguish a band-versus-Mott scenario for the 
pressure-induced insulator-metal transition: in the band-insulator-to-metal 
transition, there is no reason why, e.g, $\rho_{dc}(T)$ should show the above 
form, since neither orbital selectivity nor local antiferromagnetic spin 
fluctuations are operative there. More detailed transport work to corroborate 
our prediction are thus called for in future.
 
\vspace{0.25cm}
\section{Discussion and Conclusion}

Insofar as superconductivity arises at the boundary of the Mott transition, 
our analysis  provides tantalizing insight into sources of the pairing glue. 
Since the incoherent metal has a finite residual entropy ($S\propto ln~2$ 
per site) from the Mott localized $p_{x}$ sector, this electronic system is 
inherently unstable to soft two-particle instabilities~\cite{ourSC}. In 
solid O$_{2}$ at high pressure, lack of conditions supporting magnetic and 
charge-density instabilities in the correlated electronic structure near the 
insulator-metal transition opens the door to superconductivity as the only 
two-particle instability that can quench this finite normal state entropy. 
In fact, the situation is quite similar to that considered by Capone 
{\it et al.}~\cite{capone} in the fulleride context, where the pseudogapped, 
bad metal arose from an unstable (intermediate coupling) fixed point in the 
impurity problem, corresponding to the Kondo unscreened phase: in our case, 
precisely the same effect results from selective localization of the $p_{x}$ 
band at the orbital-selective Mott phase. In fact, in the orbital-selective 
metal, the low-energy physics is selfconsistently controlled by strong 
scattering between quasi-itinerant $p_{y,z}$ and Mott localized $p_{x}$ orbital 
states, implying low-energy singularities in one- and two-particle
propagators~\cite{pwa}. This suggests soft, multi-orbital electronic modes
at low energy, which can potentially act as a pair glue.  In a way similar 
to the fulleride case, we then expect that multiband spin-singlet $s$-wave
superconductivity (notice that $J_{H}<<U$, favoring $S=0$), driven by
such soft inter-orbital electronic fluctuations in this unstable
phase, will cut off the incoherent metal found above, and that the
superconducting transition temperature $T_{c}$ will rise to  values larger 
than those obtained for the weakly correlated case~\cite{capone}. 
Interestingly, the variation of $T_{c}$ with decreasing $U/W$ (increasing 
pressure) found by Capone {\it et al.} does bear uncanny resemblance 
(Fig.~4 of ref.~\onlinecite{capone}) to the $T_{c}(P)$ observed in solid 
O$_{2}$ under high pressure. This is suggestive, but out of scope of the 
present work. We leave details for the future.

In conclusion, we have theoretically studied the insulator-metal transition in 
highly pressurized solid O$_{2}$ using first-principles 
local-density-approximation plus dynamical-mean-field calculations. In analogy 
with multi-orbital $d$- and $f$-band systems, we find an orbital-selective 
Mott transition and an incoherent metallic normal state, arising from the 
Mott insulator via a weakly first-order transition as a function of pressure. 
Implications of our picture for the superconducting state are discussed: we 
propose that soft, multi-orbital electronic fluctuations involving 
dualistic states, i.e, the quasi-itinerant ($p_{y},p_{z}$) and Mott 
localized ($p_{x}$) states arising at this orbital-selective Mott transition 
act as the pairing glue for the superconducting state found at low $T$ in 
solid O$_{2}$. Our work underlines the importance of local dynamical 
correlations in this molecular-solid system, and holds promise for 
understanding similar physics in other solidified gases.

\vspace{0.25cm}
\hspace{-0.5cm}
{\bf Methods} \newline
To reveil the electronic reconstruction at the border of the Mott 
metal-insulator transition in solid Oxygen, we employ an state-of-the-art 
implementation of LDA+DMFT, which correctly takes disorder, temperature 
and pressure effects into account, in multi-band systems~\cite{v2o3}. 
The one-particle, LDA density-of-states are computed using the non-fully 
relativistic version of the PY-LMTO code~\cite{claudia}. To incorporate 
the effects of dynamical electronic correlations in solid O$_2$, we use 
the multi-orbital iterated-perturbation-theory (MO-IPT) as an impurity 
solver of the many-particle problem in DMFT~\cite{ePAM}. Finally, we 
carried out the computation of electrical transport within the Kubo 
formalism~\cite{class}.

\mysection{Acknowledgements}
\noindent The authors acknowledge interesting discussions with E. Tosatti
at the early stages of this work. This work was supported by CNPq 
(Proc. No. 307487/2014-8) and DFG SPP 1415. S.L. acknowledges ZIH Dresden 
for the generous allocation of computational time. L.C. thanks the 
Institut f\"ur Theoretische Chemie, Technische Universit\"at Dresden, 
for hospitality. S.L. wishes to thank the DFG for a personal Heisenberg 
Grant (Heisenberg Program).

\mysection{Author contributions}
\noindent S.L. performed {\it ab initio} (LDA) calculations. L.C. conceived 
the project and performed LDA+DMFT calculations. M.S.L. and L.C. analyzed the 
LDA+DMFT results and wrote the manuscript. All authors discussed the results 
and reviewed the manuscript.

\mysection{Additional information}
\noindent {\bf Competing financial interests:} The authors declare no 
competing financial interests.\\
\noindent Correspondence and requets for materials should be addressed 
to L.C. (lcraco@fisica.uftm.br).

\end{document}